\documentclass[12pt,fleqn,draft]{article}
\usepackage{mcite,amsfonts,style}
\begin{document}
\begin{titlepage}
\pub{4}{96}
\title{$\calO$(\alphastwo) Contributions to the longitudinal fragmentation
      function in $e^+\,e^-$ annihilation}
    {P.J. Rijken and W.L. van Neerven}
    {March 1996}
\abstract{
We present the order \alphastwo~contributions to the coefficient functions
corresponding to the longitudinal fragmentation function $F_L(x,Q^2)$. A comparison
with the leading order \alphas~result for $F_L(x,Q^2)$ shows that the corrections
are large and vary from 44\% to 67\% in the region $0.01 < x < 0.9$ at $Q^2=M_Z^2$.
Our
calculations also reveal that the ratio of the longitudinal and total cross section
$\sigma_L/\sigma_{\rm tot}$ amounts to 0.054. This number is very close to the
most recent
value obtained by the OPAL collaboration which obtained $0.057\pm 0.005$.
}
\end{titlepage}
%

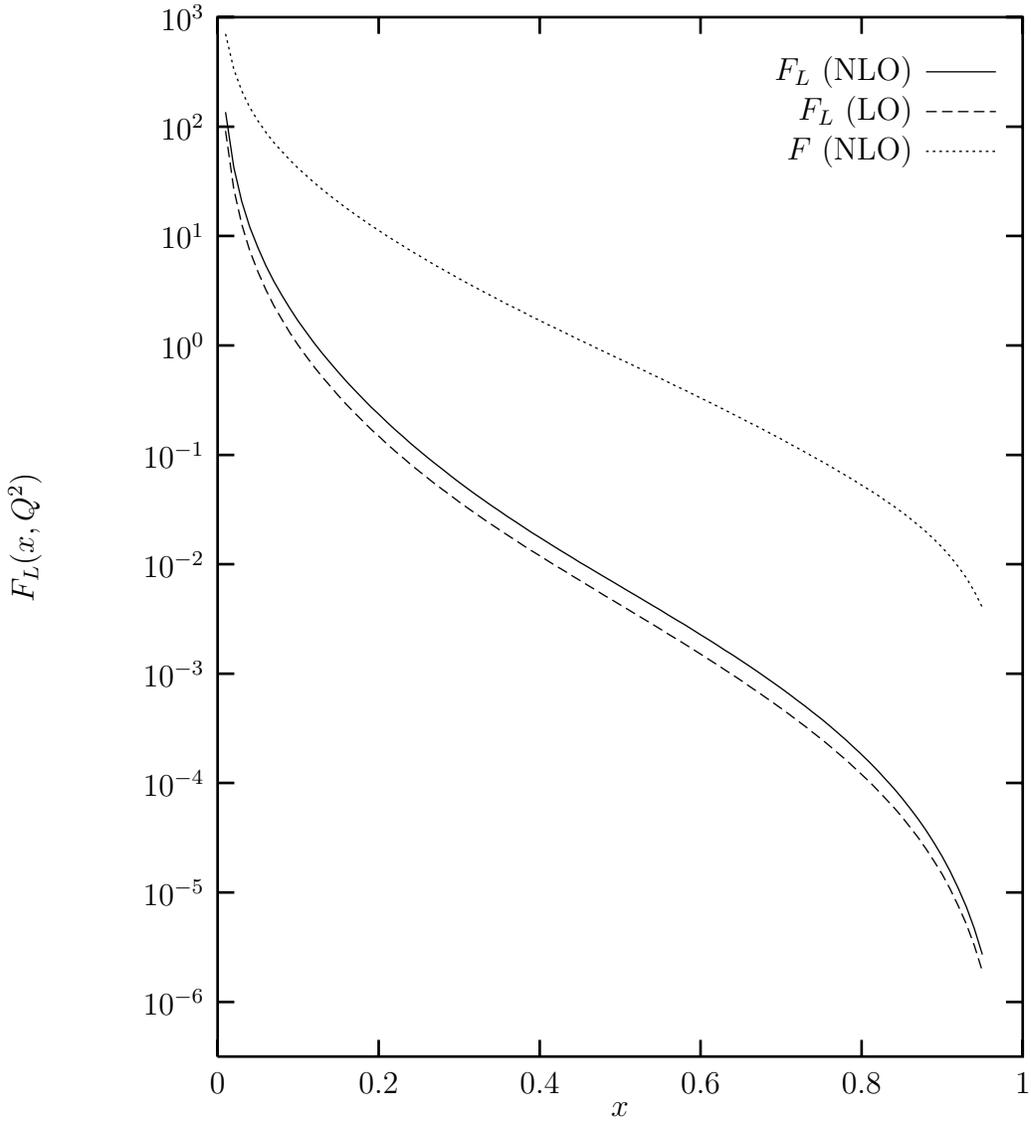
\begin{figure}[htbp]
  \input pic1
  \caption{\label{fig:fl}}
\end{figure}

\newpage
In this paper we present the $\calO$(\alphastwo) QCD corrections to the
longitudinal fragmentation function measured in the process
\begin{equation}
  \label{eq:process}
  e^+\,e^- \rightarrow \gamma,Z \rightarrow H + ``X",
\end{equation}
where $``X"$ denotes any inclusive final hadronic state and $H$ represents either a
specific charged outgoing hadron or a sum over all charged hadron species.
This process has been studied over a wide range of energies at many different
$e^+\,e^-$
colliders. For the most recent experimental results we refer to \cite{Ake95,Bus95}.
Following
the notations in \cite{Nas94} the unpolarized differential cross section of process
\eref{eq:process} is given by
\begin{equation}
  \label{eq:diff_cross}
  \frac{d^2\sigma}{dx\,d\cos\theta} = \frac{3}{8}(1+\cos^2\theta)\,
  \frac{d\sigma_T}{dx}
  + \frac{3}{4}\sin^2\theta\,\frac{d\sigma_L}{dx}
  + \frac{3}{4}\cos\theta\,\frac{d\sigma_A}{dx}.
\end{equation}
The Bj{\o}rken scaling variable $x$ is defined by
\begin{equation}
  \label{eq:bjorken}
  x = \frac{2pq}{Q^2} \hspace*{2in},\hspace*{4mm} 0 < x \leq 1,
\end{equation}
where $p$ and $q$ ($q^2 = Q^2 > 0$) are the four-momenta of the produced particle
$H$ and the
virtual vector boson ($\gamma$, $Z$) respectively. The variable $\theta$ denotes
the angle of
emission of particle $H$ with respect to the electron beam direction in the CM frame.
The transverse, longitudinal and asymmetric cross sections in \eref{eq:diff_cross}
are
defined by $\sigma_T$, $\sigma_L$ and $\sigma_A$ respectively.
The latter
only shows up if the intermediate vector boson is given by the $Z$-boson and is
absent
in purely electromagnetic annihilation. From the cross sections $\sigma_k$ ($k =
T, L, A$) one infers the fragmentation functions $F_k(x,Q^2)$ which are
the analogues of the deep inelastic structure functions measured in deep inelastic
lepton-hadron scattering where $q$ is spacelike. The former are defined by
\begin{equation}
  \label{eq:structure}
  \frac{d\sigma_k}{dx} = \sigma_k^{(0)}(Q^2)\,F_k(x,Q^2).
\end{equation}
$\sigma_k^{(0)}(Q^2)$ stands for the pointlike annihilation cross section presented
in
eq. 2.12 of \cite{Nas94}. In the framework of the QCD improved parton model the
fragmentation functions $F_k(x,Q^2)$ can be expressed as follows
\begin{equation}
  \label{eq:fragment}
  F_k(x,Q^2) = \int_x^1\frac{dz}{z}\left[ D_q\left(\frac{x}{z},\mu^2\right)\,
  \mathbb{C}_{k,q}\left(z,\frac{Q^2}{\mu^2}\right) + D_g\left(\frac{x}{z},\mu^2
  \right)\,
  \mathbb{C}_{k,g}\left(z,\frac{Q^2}{\mu^2}\right)\right],
\end{equation}
where $D_q$ and $D_g$ are the light quark and gluon fragmentation functions as
defined
in \cite{Nas94} where one has summed over all charged hadron species. Furthermore
the timelike light quark and gluon coefficient functions are given by
$\mathbb{C}_{k,q}$
and $\mathbb{C}_{k,g}$ respectively where $\mu$ denotes the factorization scale which
for convenience has been set equal to the renormalization scale. Notice that one
can also
include in \eref{eq:fragment} the contributions coming from the heavy quark coefficient
functions
and higher twist effects as has been done in \cite{Nas94}.\\
The above coefficient functions have been calculated up to first order in the strong
coupling constant $\alpha_s(\mu^2)$ in \cite{Alt79b,Bai79} where they are presented
in the \MS-scheme. Using the next-to-leading order timelike DGLAP splitting
functions,
which have been computed in \cite{Cur80,*Fur80,*Flo81}, one has made a complete
next to leading order (NLO) analysis for $F_T(x,Q^2)$ in \cite{Nas94}. This
analysis has been performed in the annihilation scheme in \cite{Nas94} where one has
also included the contributions due to the charm and bottom quarks and higher twist
effects. Here the annihilation scheme (AS) is defined by requiring that the total
fragmentation function
\begin{equation}
  \label{eq:tot_frag}
  F(x,Q^2) = F_T(x,Q^2) + F_L(x,Q^2),
\end{equation}
does not recieve QCD corrections via the coefficient functions provided $\mu^2 = Q^2$.
However the NLO analysis of the longitudinal fragmentation function $F_L(x,Q^2)$
is not complete yet because the $\calO$(\alphas) contributions to its coefficient
functions are scheme independent so that the scheme dependence of the two-loop DGLAP
splitting functions is not cancelled up to $\calO$(\alphastwo). We want to fill in
this gap by including the $\calO$(\alphastwo) contributions to the longitudinal
coefficient functions $\mathbb{C}_{L,i}(z,Q^2/\mu^2)$ ($i = q,g$) so that one
can study the effects of the latter on the analysis of $F_L(x,Q^2)$ in
\cite{Nas94}.\\
The computation of $\mathbb{C}_{k,i}$ also leads to the $\calO$(\alphastwo)
contributions to the transverse and longitudinal total cross sections defined by
\begin{equation}
  \label{eq:cross_parts}
  \sigma_k(Q^2) = \frac{1}{2}\int_0^1 dx\,x\,\frac{d\sigma_k(x,Q^2)}{dx} =
  \frac{1}{2}\,\sigma_k^{(0)}(Q^2)\,\int_0^1 dx\,x\,F_k(x,Q^2),
\end{equation}
from which one can derive the total cross section
\begin{equation}
  \label{eq:tot_cross}
  \sigma_{\rm tot}(Q^2) = \sigma_T(Q^2) + \sigma_L(Q^2),
\end{equation}
and the ratios $\sigma_k/\sigma_{\rm tot}$. Before the calculations presented below
these ratios are only known up to $\calO$(\alphas) and one obtains
\begin{equation}
  \label{eq:ratios}
  \frac{\sigma_T}{\sigma_{\rm tot}} = 1 - \frac{\alpha_s}{\pi}
  \hspace*{1in},\hspace*{4mm}
  \frac{\sigma_L}{\sigma_{\rm tot}} = \frac{\alpha_s}{\pi}.
\end{equation}
Choosing \alphas($M_Z$) = 0.126 \cite{Bus95} the quantities $\sigma_T/\sigma_{\rm tot}$ and
$\sigma_L/\sigma_{\rm tot}$ become 0.960 and 0.040 respectively. These numbers have to be
compared with the most recent experimental values obtained in \cite{Ake95} which
yield $\sigma_T/\sigma_{\rm tot} = 0.943\pm 0.005$ and
$\sigma_L/\sigma_{\rm tot} = 0.057\pm 0.005$. In particular the experimental
value of $\sigma_L/\sigma_{\rm tot}$ is far above the theoretical $\calO$(\alphas)
prediction which indicates that higher order QCD corrections are important.\\
The calculation of the $\calO$(\alphastwo) longitudinal coefficient functions proceeds
in exactly the same way as done for the Drell-Yan process in \cite{Ham91,*Nee92} and
deep inelastic lepton hadron scattering in \cite{Zij92}. First one computes the parton
fragmentation functions $\hat{\cal F}_{L,i}$ ($i=q,g$) corresponding to the process
\begin{equation}
  \label{eq:parton}
  V \rightarrow ``a" + a_1 + a_2 + \ldots + a_n,
\end{equation}
where $V = \gamma, Z$, $``a"$ denotes the detected parton and $a_i$ ($i = 1\ldots n$)
stand for the partons of which the momenta are integrated over so that the process
is inclusive with respect to the $a_i$. In zeroth order of \alphas~we have the Born
reaction
\begin{equation}
  \label{eq:born}
  V \rightarrow ``q" + \bar{q},
\end{equation}
which contributes to $\hat{\cal F}_{T,q}$ and $\hat{\cal F}_{A,q}$ but not to
$\hat{\cal F}_{L,q}$. In next-to-leading (NLO) order one obtains the one-loop virtual
corrections to reaction \eref{eq:born} and the parton subprocesses:
\begin{eqnarray}
  \label{eq:parton_sub_q}
  V & \rightarrow & ``q" + \bar{q} + g, \\[2ex]
  \label{eq:parton_sub_g}
  V & \rightarrow & ``g" + q + \bar{q}.
\end{eqnarray}
Besides $\hat{\cal F}_{T,i}$, $\hat{\cal F}_{A,i}$ ($i=q,g$), the longitudinal
fragmentation function $\hat{\cal F}_{L,i}$  now also receives contributions from the
above subprocesses.\\
After mass factorization the collinear divergences are removed and the objects
$\hat{\cal F}_{k,i}$ turn into the coefficient function which are presented in \cite{Bus95},
\cite{Alt79b} and \cite{Bai79}. The determination of the $\calO$(\alphastwo) corrections
involves the computation of the two-loop corrections to \eref{eq:born} and the one-loop
corrections to \eref{eq:parton_sub_q}, \eref{eq:parton_sub_g}. Furthermore one has to
include the following subprocesses:
\begin{eqnarray}
  \label{eq:parton_sub_qqgg}
  V & \rightarrow & ``q" + \bar{q} + g + g,\\[2ex]
  \label{eq:parton_sub_gqqg}
  V & \rightarrow & ``g" + q + \bar{q} + g,\\[2ex]
  \label{eq:parton_sub_qqqq}
  V & \rightarrow & ``q" + \bar{q} + q + \bar{q}.
\end{eqnarray}
In reaction \eref{eq:parton_sub_qqqq} the two anti-quarks can be  identical
as well as
non identical. Notice that in the above reactions the detected quark can be replaced
by the detected anti-quark so that in reaction \eref{eq:parton_sub_qqqq} one can also
distinguish between final states containing identical quarks and non identical
quarks.
In this paper we are only interested in the computation of
$\hat{\cal F}_{L,i}$.
Since the latter does not depend on the intermediate vector boson we perform our
calculation for $V=\gamma$. Notice that this also holds for
$\hat{\cal F}_{T,i}$.
After mass factorization and renormalization for which we have chosen the \MS-scheme
(indicated by a bar on the coefficient functions) the longitudinal coefficient
functions read as follows. For the non-singlet part we have up to $\calO$(\alphastwo)
\par
\begin{eqnarray}
  \label{eq:clqns}
  \lefteqn{
\overline{\mathbb{C}}_{L,q}^{NS}(x,Q^2/\mu^2) =
\frac{\alpha_s}{4\pi}\,C_F\,[2] + \left(\frac{\alpha_s}{4\pi}\right)^2\BigLeftHook
C_F^2\,\BigLeftBrace [ 8\ln(1-x) - 4\ln x + 2 + 4 x ]\cdot} \nonumber\\[2ex]
&& \cdot\ln\frac{Q^2}{\mu^2}
+ 16 S_{1,2}(1-x) + 32 S_{1,2}(-x) - 48 \Li_3(-x) + 32 \ln(1+x)\,\Li_2(-x) \nonumber\\[2ex]
&& \mbox{}
+ 16 \ln x\,\Li_2(-x) + 16 \zeta(2)\,\ln(1+x) - 16 \zeta(2) \ln(1-x)
+ 16 \ln x\,\ln^2(1+x) \nonumber\\[2ex]
&& \mbox{}- 8\ln^2 x\,\ln(1+x) - 16 \zeta(3)
+ \left(- 16 + \frac{48}{5} x^{-2} - 32 x - \frac{32}{5} x^3\right) (\Li_2(-x)\nonumber\\[2ex]
&& \mbox{}+ \ln x\,\ln(1+x) ) - 20\Li_2(1-x) + 4\ln x\,\ln(1-x)
+ \BigLeftParen 16 - 32 x - \frac{32}{5} x^3 \BigRightParen\cdot\nonumber\\[2ex]
&& \cdot\zeta(2) + 4\ln^2(1-x)
+ \left( - 6 + 16 x + \frac{16}{5} x^3\right) \ln^2 x
+ ( 14 + 4 x )\,\ln(1-x)\nonumber\\[2ex]
&& \mbox{}+ \left(\frac{74}{5} - \frac{48}{5} x^{-1} + \frac{64}{5} x
+ \frac{32}{5} x^2\right)\,\ln x
- \frac{27}{5} + \frac{48}{5} x^{-1} - \frac{138}{5} x + \frac{32}{5} x^2\BigRightBrace
\nonumber\\[2ex]
&& \mbox{}+ C_A\,C_F\BigLeftBrace  - \frac{22}{3}\,\ln\frac{Q^2}{\mu^2} - 8 S_{1,2}(1-x)
- 16 S_{1,2}(-x) + 24 \Li_3(-x) \nonumber\\[2ex]
&& \mbox{}- 16 \ln(1+x)\,\Li_2(-x) - 8 \zeta(2)\,\ln(1+x)
+ 8 \zeta(2)\,\ln(1-x) - 8 \ln x\,\Li_2(-x) \nonumber\\[2ex]
&& \mbox{}- 8 \ln x\,\ln^2(1+x)
+ 4 \ln^2 x\,\ln(1+x) + 8 \zeta(3)
+ \BigLeftParen8 - \frac{24}{5} x^{-2} + 16 x \nonumber\\[2ex]
&& \mbox{}+ \frac{16}{5} x^3\BigRightParen (\Li_2(-x) + \ln x\ln(1+x) ) + 4\Li_2(1-x)
+ \zeta(2)\left(16 x + \frac{16}{5} x^3\right) \nonumber\\[2ex]
&& \mbox{}- \left(8 x + \frac{8}{5} x^3\right)
\ln^2 x - \frac{46}{3}\,\ln(1-x) + \left(- \frac{206}{15} + \frac{24}{5} x^{-1}
- \frac{12}{5} x - \frac{16}{5} x^2\right)\,\ln x \nonumber\\[2ex]
&& \mbox{}+ \frac{1189}{45} - \frac{24}{5} x^{-1} + \frac{82}{15} x - \frac{16}{5} x^2
\BigRightBrace + n_f\,C_F\,T_f\,\BigLeftBrace \frac{8}{3}\,\ln\frac{Q^2}{\mu^2}
+ \frac{8}{3}\,(\ln(1-x) \nonumber\\[2ex]
&& \mbox{}+ \ln x) - \frac{100}{9} + \frac{8}{3} x\BigRightBrace
+ \left(C_F^2 - \frac{1}{2}C_A\,C_F\right)\BigLeftBrace
8\Li_2(1-x) - 8(1+x)\,\ln x \nonumber\\[2ex]
&& \mbox{}- 24 (1-x)\BigRightBrace\BigRightHook,
%

\end{eqnarray}
where the colour factors are given by $C_F = (N^2-1)/2N$, $C_A = N$ and
$T_f = 1/2$ ($N=3$ in QCD). The quantity $n_f$ denotes the number of light flavours.
The definitions for the polylogarithmic functions $\Li_2(x)$ and $S_{n,p}(x)$
can be found in \cite{Lew83,*Bar72,*Dev84}.
In
NLO the terms proprtional to $C_F^2$, $C_A\,C_F$ and
$C_F\,T_f$
receive contributions from identical as well as non identical (anti) quarks in reaction
\eref{eq:parton_sub_qqqq} whereas the term proportional to $C_F^2-C_A\,C_F/2$
can be only attributed to identical (anti) quarks. The singlet coefficient function can
be written as
\begin{equation}
  \label{eq:singlet}
  \overline{\mathbb{C}}_{L,q}^S(x,Q^2/\mu^2) =
  \overline{\mathbb{C}}_{L,q}^{NS}(x,Q^2/\mu^2) +
  \overline{\mathbb{C}}_{L,q}^{PS}(x,Q^2/\mu^2).
\end{equation}
The pure singlet part denoted by $\mathbb{C}_{L,q}^{PS}$ originates from process
\eref{eq:parton_sub_qqqq} where one gluon is exchanged in the $t$-channel. It reads as
follows
\begin{eqnarray}
  \label{eq:clqps}
  \lefteqn{
\overline{\mathbb{C}}_{L,q}^{PS}(x,Q^2/\mu^2) =
\left(\frac{\alpha_s}{4\pi}\right)^2\,n_f\,C_F\,T_f\,\BigLeftHook
\BigLeftBrace 16\ln x + \frac{32}{3} x^{-1} - 16 x + \frac{16}{3} x^2
\BigRightBrace\ln\frac{Q^2}{\mu^2}} \nonumber\\[2ex]
&& \mbox{}+ 16\Li_2(1-x) + 16\ln x\,\ln(1-x)
+ 24\ln^2 x + \left(\frac{32}{3} x^{-1} - 16 x + \frac{16}{3} x^2\right)\cdot \nonumber\\[2ex]
&& \cdot\ln(1-x)
+ \left(- 32 + \frac{64}{3} x^{-1} - 32 x + \frac{16}{3} x^2\right) \ln x
- \frac{112}{3} - 16 x^{-1} + \frac{208}{3} x \nonumber\\[2ex]
&& \mbox{}- 16 x^2 \BigRightHook.
%

\end{eqnarray}
Finally we have the gluonic coefficient function which is due to the parton subprocesses
\eref{eq:parton_sub_g},\eref{eq:parton_sub_gqqg}. It becomes equal to
\begin{eqnarray}
  \label{eq:clg}
  \lefteqn{
\overline{\mathbb{C}}_{L,g}(x,Q^2/\mu^2) = \frac{\alpha_s}{4\pi}\,C_F
\left[ \frac{8}{x}-8\right] + \left(\frac{\alpha_s}{4\pi}\right)^2\BigLeftHook
C_F^2\BigLeftBrace\BigLeftHook 16\ln x - 8 + 16 x^{-1} - 8 x\BigRightHook
\cdot}\nonumber\\[2ex]
&& \cdot\ln\frac{Q^2}{\mu^2} + \left(- \frac{32}{3} + \frac{64}{5} x^{-2}
+ \frac{32}{15} x^3\right) (\Li_2(-x) + \ln x\ln(1+x) ) + 16 \Li_2(1-x)
\nonumber\\[2ex]
&& \mbox{}+ 16 \ln x\,\ln(1-x) + \frac{32}{15}\zeta(2) x^3
+ \left(24 - \frac{16}{15} x^3\right)\,\ln^2 x
+ \left(- 24 + 32 x^{-1} - 8 x\right)\cdot \nonumber\\[2ex]
&& \cdot\ln(1-x) + \left(- \frac{8}{5} + \frac{96}{5} x^{-1} - \frac{224}{15} x
- \frac{32}{15} x^2\right)\,\ln x
+ \frac{24}{5} - \frac{96}{5} x^{-1} + \frac{248}{15} x \nonumber\\[2ex]
&& \mbox{}- \frac{32}{15} x^2\BigRightBrace + C_A\,C_F\,\BigLeftBrace
\BigLeftHook \left( - 32 + 32 x^{-1}\right) \ln(1-x) - \left(32 + 32 x^{-1}\right)\ln x + 80
\nonumber\\[2ex]
&& \mbox{}- \frac{272}{3} x^{-1} + 16 x - \frac{16}{3} x^2\BigRightHook\,\ln\frac{Q^2}{\mu^2}
+ \left(32 + 32 x^{-1}\right) (\Li_2(-x) + \ln x\ln(1+x)) \nonumber\\[2ex]
&& \mbox{}- \frac{64}{x}\Li_2(1-x)
- 64\ln x\ln(1-x)
+ \zeta(2) \left(- 64 + 96 x^{-1}\right) + \left(- 16 + 16 x^{-1}\right)\cdot
\nonumber\\[2ex]
&& \cdot\ln^2(1-x) - \left(48 + 64 x^{-1}\right)\,\ln^2 x
+ \left(144 - \frac{464}{3} x^{-1} + 16 x - \frac{16}{3} x^2\right)\,\ln(1-x)
\nonumber\\[2ex]
&& \mbox{}+ \left(112 - \frac{352}{3} x^{-1} + 32 x - \frac{16}{3} x^2\right)\,\ln x
- \frac{320}{3} + \frac{448}{3} x^{-1} - \frac{160}{3} x + \frac{32}{3} x^2
\BigRightBrace\BigRightHook. \nonumber\\[2ex]
&&
%

\end{eqnarray}
Besides in the \MS-scheme the mass factorization can also be performed in the AS-scheme
(see above \eref{eq:tot_frag}). Using the transformation formulae in \cite{Nas94} we
obtain
\begin{eqnarray}
  \label{eq:clqns_as}
  \lefteqn{
    \mathbb{C}_{L,q}^{NS}(x,Q^2/\mu^2) =
    \overline{\mathbb{C}}_{L,q}^{NS}(x,Q^2/\mu^2) + \left(\frac{\alpha_s}{4\pi}\right)^2
    \,C_F^2 \BigLeftHook 20 \Li_2(1-x)} \nonumber \\[2ex]
  && + 4\ln x\,\ln(1-x) - 16\zeta(2) - 4\ln^2(1-x) + 4\ln^2 x + (10-4x)\,\ln(1-x)
  \nonumber \\[2ex]
  && + (4-8x)\,\ln x+ 18 + 6x \BigRightHook,\\[2ex]
  \label{eq:clqps_as}
  \lefteqn{
    \mathbb{C}_{L,q}^{PS}(x,Q^2/\mu^2) =
    \overline{\mathbb{C}}_{L,q}^{PS}(x,Q^2/\mu^2)} ,\\[2ex]
  \label{eq:clg_as}
  \lefteqn{
    \mathbb{C}_{L,g}(x,Q^2/\mu^2) =
    \overline{\mathbb{C}}_{L,g}(x,Q^2/\mu^2) + \left(\frac{\alpha_s}{4\pi}\right)^2
    \,C_F^2 \BigLeftHook -16 \Li_2(1-x)} \nonumber\\[2ex]
  && - 16 \ln x\,\ln(1-x) - 16 \ln^2 x- \left(\frac{16}{x} - 8 - 8 x\right)\,\ln(1-x)
  \nonumber \\[2ex]
  && - \left(\frac{32}{x} + 16 - 16 x\right)\,\ln x - \frac{32}{x} + 56 - 24 x
  \BigRightHook.
\end{eqnarray}
Notice that in the above formulae we did not distinguish between the renormalization
and factorization scale. The distinction can be easily made if one substitutes in the
above coefficient functions
\begin{equation}
  \label{eq:renorm_fac}
  \alpha_s(\mu^2) = \alpha_s(\mu_R^2)\,\left[1 + \frac{\alpha_s(\mu_R^2)}{4\pi}
  \left( \frac{11}{3}C_A - \frac{4}{3}n_f\,T_f\right)\,\ln\frac{\mu_R^2}{\mu^2}
    \right],
\end{equation}
where $\mu$ and $\mu_R$ denote the factorization and renormalization scale respectively.
Inspection of the above coefficient functions reveals that there are logarithms of the
type $\ln^kx/x$ ($k=0,1,2$) present in the coefficient functions
$\overline{\mathbb{C}}_{L,q}^{PS}$ \eref{eq:clqps} and
$\overline{\mathbb{C}}_{L,g}^{PS}$ \eref{eq:clg} which become large when
$x\rightarrow 0$. The relevance of these terms for the NLO corrections to $F_L
(x,Q^2)$ \eref{eq:fragment} will be discussed at the end of the paper. The ratios
$\sigma_k/\sigma_{\rm tot}$ ($k = T, L$) are given by
\begin{equation}
  \label{eq:ratios_cross}
  \frac{\sigma_k}{\sigma_{\rm tot}} = R_{e^+\,e^-}^{-1}\,\int_0^1\,dz\,z\BigLeftHook
  \,\mathbb{C}_{k,q}\left(z,\frac{Q^2}{\mu^2}\right) + \frac{1}{2}\,
  \mathbb{C}_{k,g}\left(z,\frac{Q^2}{\mu^2}\right)\BigRightHook,
\end{equation}
where $R_{e^+\,e^-}$ is defined by $R_{e^+\,e^-} = \sigma_{\rm tot}/\sigma^{(0)}$.
Furthermore we have used in the derivation of \eref{eq:ratios_cross} the equality
$\mathbb{C}_{k,q} = \mathbb{C}_{k,\bar{q}}$. The ratio $R_{e^+\,e^-} =
\sigma_{\rm tot}/\sigma^{(0)}$ has been calculated up to $\calO(\alpha_s^3)$ in the
literature \cite{Che79,*Din79,*Cel80,*Gor91,*Sur91}. Up to $\calO$(\alphastwo) it
is given by
\begin{eqnarray}
  \label{eq:R}
  \lefteqn{
  R_{e^+\,e^-} = 1 + \frac{\alpha_s}{4\pi}\,C_F\,[3] + \left(\frac{\alpha_s}{4\pi}\right)^2
  \,\BigLeftHook C_F^2\left\{-\frac{3}{2}\right\} + C_A\,C_F\,\BigLeftBrace
  -11\ln\frac{Q^2}{\mu_R^2} - 44\zeta(3)} \nonumber \\[2ex]
&&+\frac{123}{2}\BigRightBrace
  + n_f\,C_F\,T_f\BigLeftBrace 4\ln\frac{Q^2}{\mu_R^2} + 16\zeta(3)-22\BigRightBrace
  \BigRightHook.
\end{eqnarray}
Substituting the coefficient functions \eref{eq:clqns}-\eref{eq:clg} or
\eref{eq:clqns_as}-\eref{eq:clg_as} into \eref{eq:ratios_cross} and expanding
$R_{e^+\,e^-}^{-1}$ \eref{eq:R} up to $\calO$(\alphastwo) we obtain
\begin{eqnarray}
  \label{eq:ratio_L}
  \lefteqn{
  \frac{\sigma_L}{\sigma_{\rm tot}} = \left(\frac{\alpha_s}{4\pi}\right)\,
  C_F [3] + \left(\frac{\alpha_s}{4\pi}\right)^2\,\BigLeftHook C_F^2\,\left\{
  -\frac{33}{2}\right\} + C_A\,C_F\,\BigLeftBrace -11\ln\frac{Q^2}{\mu_R^2} -
  \frac{24}{5}\zeta(3)} \nonumber \\[2ex]
  && + \frac{2023}{30}\BigRightBrace
  + n_f\,C_F\,T_f\BigLeftBrace 4\ln\frac{Q^2}{\mu_R^2}-\frac{74}{3}
  \BigRightBrace\BigRightHook, \\[2ex]
  \label{eq:ratio_T}
  \lefteqn{
  \frac{\sigma_T}{\sigma_{\rm tot}} = 1 - \frac{\sigma_L}{\sigma_{\rm tot}}.}
\end{eqnarray}
Notice that the above perturbation series are factorization-scheme independent.
They do however
depend on the renormalization scheme (here \MS) which is indicated by $\mu_R$. Since
$\sigma_{\rm tot} = \sigma^{(0)}\,R_{e^+\,e^-}$ we infer that in zeroth order,
$\sigma_{\rm tot}$ only receives contributions via $\sigma_T$ whereas in first order
of \alphas, $\sigma_{\rm tot}$ is determined by $\sigma_L$. In second order,
$\sigma_L$ as well as $\sigma_T$ contribute to $\sigma_{\rm tot}$.\\
Choosing $\alpha_s(M_Z)$ = 0.126 and $n_f = 5$ \cite{Bus95} we get the following results
\begin{eqnarray}
  \label{eq:result_L}
  \frac{\sigma_L}{\sigma_{\rm tot}} &=& 0.040 + 0.014 = 0.054,\\[2ex]
  \label{eq:result_T}
  \frac{\sigma_T}{\sigma_{\rm tot}} &=& 0.946,
\end{eqnarray}
which are very close to the values obtained by the OPAL-experiment \cite{Ake95}
mentioned below \eref{eq:ratios}. In \eref{eq:result_L} we have explicitely
shown the $\calO$(\alphas) and $\calO$(\alphastwo) contribution which are
represented by the numbers 0.040 and 0.014 respectively. From the latter we conclude
that the NLO correction to $\sigma_L/\sigma_{\rm tot}$ amounts to about 30\%
of the LO one. The agreement between the NLO theoretical predictions and the experimental
results show that equations \eref{eq:ratio_L} and \eref{eq:ratio_T} are very suitable
to determine the strong coupling constant provided we also include the contributions
due to heavy quarks (here $c$ and $b$) and higher twist effects (see \cite{Nas94}).\\
Finally we want to study the effect of the $\calO$(\alphastwo) corrections to the
longitudinal fragmentation function $F_L(x,Q^2)$. Adopting the AS-scheme
(see \cite{Cur80,*Fur80,*Flo81} above) and choosing the sum over the light quark
fragmentation functions $D_t = \sum_q\,D_q$ and the gluon fragmentation function $D_g$
in table 1 of \cite{Nas94} we have computed $F_L(x,Q^2)$ in \cite{Bai79} up
to NLO. Further we took $\alpha_s(M_Z)$ = 0.126 (\MS) and $n_f=5$ \cite{Bus95}.
In fig.~\ref{fig:fl}
we have plotted the LO as well the NLO result for $0.01 < x < 0.9$ and have shown for
comparison $F(x,Q^2)$ in \eref{eq:tot_frag}. Notice that the data of the experiments
in \cite{Ake95,Bus95} are taken in the range  $0.01 < x < 0.9$. To exhibit the NLO
corrections more clearly we have also plotted the ratio $K=F_L^{\rm NLO}/
F_L^{\rm LO}$ in fig.~\ref{fig:ratio}. The last figure shows that the
$\calO$(\alphastwo) corrections to $F_L(x,Q^2)$ are large in spite of the fact that
$\alpha_s(M_Z)$ is small. They amount to 45\% at $x=0.01$ and increase to 67\%
around $x=0.1$. Above $x=0.1$ the corrections decrease to 44\% near $x=0.9$. We have
also studied the corrections in the region $10^{-4} < x < 10^{-2}$. They are smaller
than in the range $0.01 < x < 0.9$ with a minimum of 27\% at $x=7\cdot 10^{-3}$. A
study of the various parts contributing to $F_L(x,Q^2)$ reveals that the gluonic
part in \eref{eq:fragment} dominates $F_L(x,Q^2)$ for $x < 0.2$ whereas the light
quark contribution becomes dominant for $x > 0.2$. This holds for the $\calO$(\alphas)
as well
as $\calO$(\alphastwo) contributions. Finally we want to make a comment on the
contribution of the logarithmic term $\ln^k x/x$ mentioned under \eref{eq:renorm_fac}.
They do not dominate the corrections when convoluted with the gluon and quark
fragmentation functions $D_g$ and $D_t$. This becomes clear if one looks at the
coefficient of the $\ln^2 x/x$ term appearing in the $C_A\,C_F$ part of
$\overline{\mathbb{C}}_{L,g}$ in \eref{eq:clg}. Although this coefficient is negative
the actual total $\calO$(\alphastwo) contribution to $F_L^{\rm NLO}(x,Q^2)$
is positive over the whole $x$-range.
%

%
\bibliographystyle{/home/rulil0/pieter/tex/style/mybib}
\bibliography{/home/rulil0/pieter/tex/physics}

\newpage
\begin{mcbibliography}{10}
\bibitem{Ake95} R. Akers \etal~(OPAL), Z. Phys. {\bf C68} (1995)
  203\bibitem{Bus95} D. Buskulic \etal~(ALEPH), Phys. Lett. {\bf 357B} (1995)
  487\bibitem{Nas94} S. Nason and B.R. Webber, Nucl. Phys. {\bf B421} (1994)
  473\bibitem{Alt79b} G. Altarelli, R.K. Ellis, G. Martinelli, S.-Y. Pi, Nucl.
  Phys. {\bf B160} (1979) 301\bibitem{Bai79} R. Baier and K. Fey, Z. Phys. {\bf
  C2} (1979) 339.\bibitem{Cur80} G. Curci, W. Furmanski and R. Petronzio, Nucl.
  Phys. {\bf 175} (1980) 27\bibitem{Fur80} W. Furmanski and R. Petronzio, Phys.
  Lett. {\bf 97B} (1980) 437\bibitem{Flo81} F.G. Floratos, C. Kounnas and R.
  Lacaze, Nucl. Phys. {\bf B192} (1981) 417\bibitem{Ham91} R. Hamberg, W.L. van
  Neerven and T. Matsuura, Nucl. Phys. {\bf B359} (1991) 343\bibitem{Nee92}
  W.L. van Neerven and E.B. Zijlstra, Nucl. Phys. {\bf B382} (1992)
  11\bibitem{Zij92} E.B. Zijlstra and W.L. van Neerven, Nucl. Phys. {\bf B383}
  (1992) 525\bibitem{Lew83} L. Lewin, ``Polylogarithms and Associated
  Functions" (North Holland, Amsterdam 1983)\bibitem{Bar72} R. Barbieri, J.A.
  Mignaco and E. Remiddi, Nuovo Cim. {\bf 11A} (1972) 824\bibitem{Dev84} A.
  Devoto and D.W. Duke, Riv. Nuovo Cim. {\bf 7-6} (1984) 1\bibitem{Che79} K.E.
  Chetyrkin, A.L. Kataev and F.V. Tkachov, Phy. Lett. {\bf 85B} (1979)
  277\bibitem{Din79} M. Dine and J. Sapirstein, Phys. Rev. Lett. {\bf 43}
  (1979) 668\bibitem{Cel80} W. Celmaster and R.J. Gonsalves, Phys. Rev. Lett.
  {\bf 44} (1980) 560\bibitem{Gor91} S.G. Gorishni, A.L. Kataev and S.A. Larin,
  Phys. Lett. {\bf 159B} (1991) 144\bibitem{Sur91} L.R. Surguladze and M.A.
  Samuel, Phys. Rev. Lett. {\bf 66} (1991) 560, Erratum Phys. Rev. Lett. {\bf
  66} (1991) 2416\end{mcbibliography}
\newpage
\section{Figure captions}
{\bf Fig. 1} The leading order (LO) and next-to-leading order (NLO) corrected
             longitudinal fragmentation function $F_{\rm L}(x,Q^2)$. Dashed line:
             $F_{\rm L}^{\rm (LO)}(x,Q^2)$; solid line:
             $F_{\rm L}^{\rm (NLO)}(x,Q^2)$;
             dotted line : $F^{\rm (NLO)}(x,Q^2)$ \eref{eq:tot_frag}.\\[2ex]
{\bf Fig. 2} The ratio $K(x,Q^2) = F_{\rm L}^{\rm (NLO)}(x,Q^2)/F_{\rm L}^{\rm (LO)}
             (x,Q^2)$.
%

\newpage
%
%
%
%
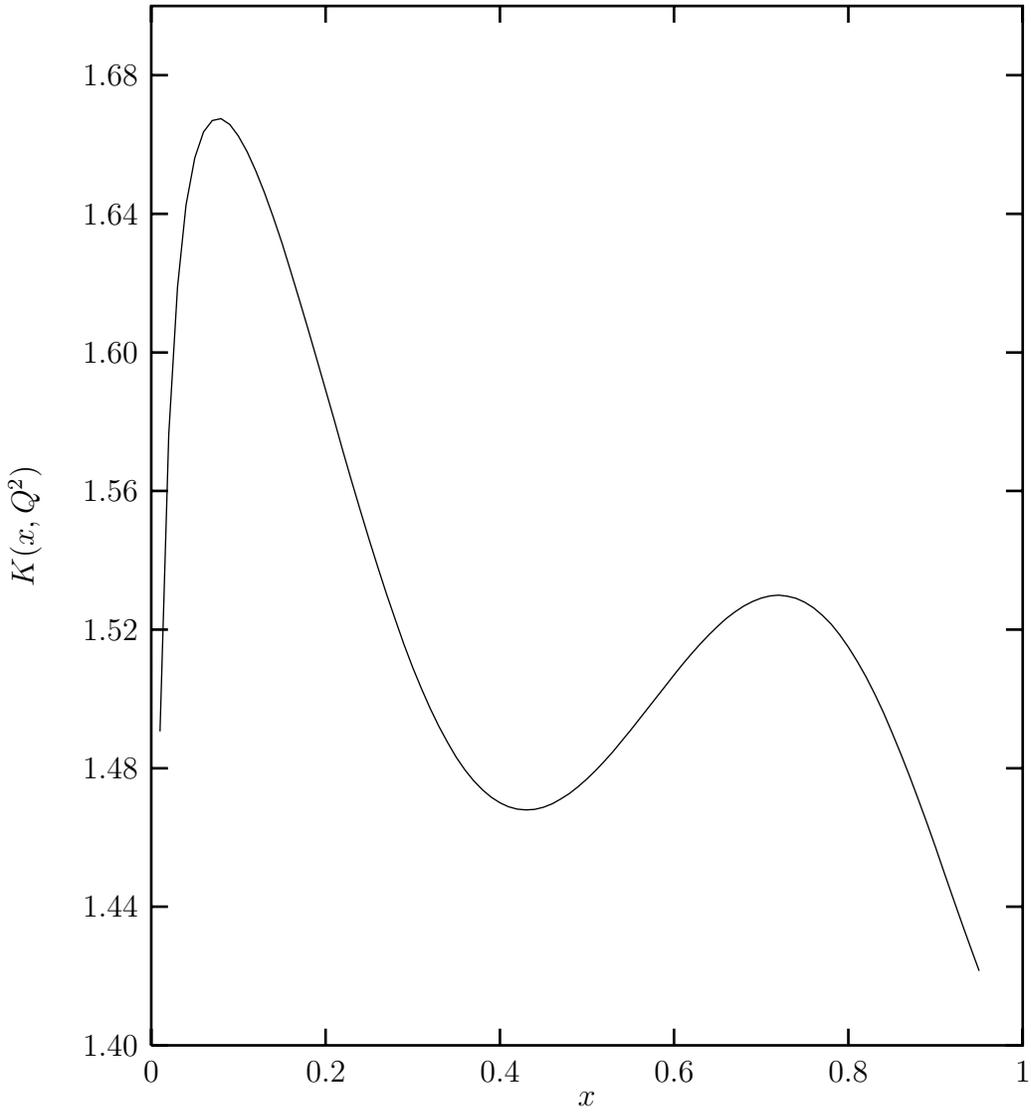
\begin{figure}[htbp]
  \input pic2
  \caption{\label{fig:ratio}}
\end{figure}
%

%
\end{document}